\documentclass{iopart}

\usepackage{graphicx}

\newcommand{\eV}{\,\mathrm{eV}}

\newcommand{\ud}{\mathrm{d}}

\def\la{\ \raise.3ex\hbox{$<$\kern-.75em\lower1ex\hbox{$\sim$}}\ }
\def\ga{\ \raise.3ex\hbox{$>$\kern-.75em\lower1ex\hbox{$\sim$}}\ }

\begin{document}
\title{Small Scale Anisotropy Predictions for the Auger Observatory}

\author{Daniel De Marco\dag, Pasquale Blasi\ddag\ and Angela V. Olinto\P}

\address{\dag\ Bartol Research Institute, University of Delaware
Newark, DE 19716, U.S.A.}

\address{\ddag\ INAF/Osservatorio Astrofisico di Arcetri,
Largo E. Fermi, 5 - 50125 Firenze, ITALY}

\address{\P\ Department of Astronomy \& Astrophysics,
Kavli Institute for Cosmological Physics,
5640 S. Ellis Ave. Chicago, IL 60637, U.S.A.}

\eads{\mailto{ddm@bartol.udel.edu}, \mailto{blasi@arcetri.astro.it},
\mailto{olinto@oddjob.uchicago.edu}}

\begin{abstract}
We study the small scale anisotropy signal expected at the Pierre
Auger Observatory in the next 1, 5, 10, and 15 years of operation,
from sources of ultra-high energy (UHE) protons. We numerically
propagate UHE protons over cosmological distances using an injection 
spectrum and normalization that fits current data up to $\sim
10^{20}\eV$. We characterize possible sources of ultra-high energy 
cosmic rays (UHECRs) by their mean density in the local Universe, 
$\bar{\rho} = 10^{-r}$ Mpc$^{-3}$, with $r$ between 3 and 6.    

These densities span a wide range of extragalactic sites for UHECR
sources, from common to rare galaxies or even clusters of galaxies.  We
simulate 100 realizations for each model and calculate the two point
correlation function for events with energies above  $4 \times
10^{19}\eV$ and above  $10^{20}\eV$, as specialized to the case of the
Auger telescope. We find that for $r\ga 4$, Auger should be able to 
detect small scale anisotropies in the near future. Distinguishing 
between different source densities based on cosmic ray data alone 
will be more challenging than detecting a departure from isotropy 
and is likely to require larger statistics of events. Combining the 
angular distribution studies with the spectral shape around the GZK 
feature will also help distinguish between different source scenarios.  
\end{abstract}

\maketitle

\section{Introduction}

There are two main open questions that together constitute the long
standing mystery of the origin of ultra-high energy
cosmic rays (UHECRs): 1) What is their acceleration site? 2) Does
their diffuse spectrum at the Earth possess the so-called GZK feature
\cite{GZK66}, evidence of their inelastic interaction with the photons 
of the cosmic microwave background (CMB)? In the next years the Pierre 
Auger Observatory \cite{auger} will provide us with important hints to 
the first question and will most likely answer the second question.
Moreover, it can be expected that this experiment will open the way to
realistic measurements of the chemical composition of cosmic rays at
the highest energies, another important piece of the puzzle. 

As the Southern site of the Auger Observatory nears completion, an 
increase by one order of magnitude of the worldwide exposure to the 
highest energy particles is within reach. The energy range between
$10^{19}$ eV and $10^{20}$ eV is of particular importance as it
provides a window for charged particle astronomy. Cosmic rays in this 
energy range suffer smaller deflections by cosmic magnetic fields than 
their lower energy counterparts, although at the lowest end of this
range the effects of the galactic magnetic field are expected to be 
important. However one can imagine to be potentially able to correct for 
this effect or even use it to infer the topology of the magnetic field 
of the Galaxy. 

A key element in determining the origin of UHECRs is the detection of
large \cite{sommers} and small scale anisotropies in their arrival 
directions. If, as
expected and as suggested by some numerical simulations (see, e.g.,
\cite{dolag04}), the deflection of cosmic rays of energies larger than 
$\sim 4 \times 10^{19}$ eV should be small, then to some extent they
should point back to their source. As large numbers of cosmic rays at 
such energies are observed, the pointing becomes more easily
feasible. The net result is the appearance of small angle clustering
in the arrival directions, and the strength of the effect is related
to the mean density of sources in the universe. 

In addition, interactions of cosmic rays with the cosmic background 
radiation limit the distance from which cosmic rays with 
energies above $\sim 10^{20}$ eV can reach the Earth to $\sim$ 50
Mpc. The matter distribution within a 50 Mpc radius from the Earth is 
rather anisotropic, thus large scale anisotropies in the UHECR arrival 
direction distribution should also be observable.  

An important piece of information, necessary to infer the origin of
UHECRs is their dominant chemical composition as a function of
energy. The present observational information about the composition at
the highest energies is rather poor and does not allow us to
draw any firm conclusions. While our calculations are carried out for
a purely proton composition at the highest energies, an appreciable
contamination of heavier nuclei may change our conclusions on the 
implications of small angle clustering to some extent. From the 
theoretical point of view, a proton-dominated or a mixed composition 
also have quite different implications on the transition of cosmic
rays from a galactic to an extragalactic origin: while a mixed
composition would imply that such transition takes place at the {\it
  ankle} \cite{Allard05,Allard05b}, a pure proton composition (or 
very light pollution from heavier nuclei) would imply that the 
transition occurs at lower energies ({\it dip} scenario) \cite{bere}. 

Our conclusions regarding small scale anisotropies should apply more 
generally with the caveat that we assume the best fit injection 
spectrum for a pure proton source, which is softer than the mixed 
composition case \cite{Allard05b} (however, see \cite{Armen05b} for 
the effect of magnetic fields in the source region and \cite{kachel}
for the effect of maximum energies depending on the source). Given 
that the relevant energy range is limited (between $ 4 \times 10^{19}$ 
eV and $10^{20}$ eV), small differences in the injection should not 
have a significant effect. 

The unknown magnitude and structure of the magnetic fields outside 
the Milky Way limit our ability to predict the precise energy
threshold for charged particle astronomy. The role of extragalactic
magnetic fields in changing the spectrum and clustering of UHECRs is
also badly constrained as different simulations give different
estimates for  the magnitude and spatial structure of these fields
(see, e.g., \cite{dolag04,sigl}). The combined effect of a mixed
composition and significant extragalactic fields may move the
threshold for detecting anisotropies to larger energies (see, e.g.,
\cite{Armen05a}). Here, we assume that extragalactic magnetic fields
can be neglected for particles with energies above $4\times 10^{19}$
eV. This assumption holds if magnetic fields in the extragalactic
medium are less than $\sim 0.1~\rm nG$ with a reversal scale of $\sim
1$ Mpc and the small scale anisotropies are evaluated on angles of
$\sim 1^{\circ}$  \cite{BD04,DBO05}. This magnitude  
field is compatible with observational bounds \cite{BBO99} and 
detailed numerical simulations such as \cite{dolag04} (however, see
\cite{sigl} for different numerical estimates).  
UHECR observations will eventually allow us to measure the strength
of the extragalactic magnetic field and get information about its
structure (see, e.g., \cite{L04,Deligny03}). In the mean time,
searches for small scale anisotropies should focus on energy
thresholds from  $4 \times 10^{19}$ eV to $10^{20}$ eV.  

The mean density of UHECR sources can be determined once small scale
anisotropies (SSAs) in the arrival directions are observed. Thus far
the only experiment to report departures from isotropy is AGASA
\cite{agasa_ssa}. The statistical significance of the clustering has
been challenged \cite{finley} as it depends on the angular scale
chosen for binning the data. Assuming the AGASA data, the number
density of sources was estimated to range from  $\sim 10^{-6} $
Mpc$^{-3}$  to  $\sim 10^{-4}$ Mpc$^{-3}$ with large uncertainties
\cite{BD04,DBO05,isola,others}. 
In Refs. \cite{BD04,DBO05}, full numerical simulations of the
propagation were performed which account for the statistical  errors
in the energy determination and the AGASA acceptance. The first study
concluded that the AGASA small scale anisotropies indicated a density
of sources 
$\sim 10^{-5} \rm Mpc^{-3}$  \cite{BD04}.  However, when taken
together with the observed AGASA spectrum, the SSAs and the GZK
feature become inconsistent \cite{DBO05}.   

Here we study the small scale anisotropy signal expected at the Pierre
Auger Observatory in the next 1, 5, 10, and 15 years. We numerically
propagate ultra high energy protons over cosmological distances and
characterize possible sources by their mean density in the local
Universe, $\bar{\rho} = 10^{-r}$ Mpc$^{-3}$, with $r$ between 3 and
6. These source densities span the relevant range of extragalactic
sites for UHECR sources, from galaxies to clusters of galaxies.   In
\S2 we review our numerical approach and present the SSA results of
100 realizations of each model.  We show the two point correlation
functions for events of energies above  $4\!\times\!10^{19}\eV$ and
above  $10^{20}\eV$, list the expected number of doublets, and discuss
the expected spectrum for different models. In \S3, we discuss the
implications of our results and conclude. 

\section{Small Scale Anisotropies and Spectra at Auger}

The propagation of UHECRs is simulated using the Monte-Carlo code 
described in \cite{DBO03}. We assume that UHECRs are protons 
injected with a power-law spectrum by extragalactic sources. The 
injection spectrum in taken to be of the form:
\begin{equation}
  F(E) \propto E^{-\gamma} \exp(-E/E_\mathrm{max}),
\end{equation}
where $\gamma$ is the spectral index and $E_\mathrm{max}$ is the maximum
injection energy at the source. We fix $\gamma=2.6$ and
$E_\mathrm{max}=10^{21.5}\eV$ since these values reproduce well the
experimental results in the lower energy high statistics region around  $\sim
10^{18.5}$ eV up to $\sim 10^{20}$ eV (see, e.g. \cite{DBO03}). We focus on
events above $4\times10^{19}\eV$, which are generated at $z\ll 1$,
therefore, source evolution is only marginally relevant \cite{BD04}. We
simulate the propagation of protons from the source to the observer by
including the photo-pion production, pair production, and adiabatic
energy losses due to the expansion of the Universe. We assume the
Universe has  matter and dark energy densities as  fractions of the
critical density given by  $\Omega_\mathrm{m}=0.3$ and
$\Omega_\Lambda=0.7$, respectively.  

Our source distribution in space is generated by a random placement of
sources for a given spatial density and particles are emitted from
randomly chosen source positions. The source redshifts are generated 
with a probability distribution proportional to 
\begin{equation}
   \frac{\ud n}{\ud z} \propto r(z)^2 \frac{\ud t}{\ud z} \ ,
\end{equation}
where   $\ud t/\ud  
z$ gives the relation between time and redshift (see, e.g.,  
the expression given in \cite{bg}) and $r(z)$ is defined as
\begin{equation}\label{eq:r}
   r=c\int^{t_0}_{t_\mathrm{g}}\frac{\ud t}{R(t)} \ ,
\end{equation}
where $t_\mathrm{g}$ is the age of the Universe when the event was
generated, $t_0$ is the present age of the Universe, and $R(t)$ is the
scale factor of the Universe. Once a particle has been generated, the
code propagates it to the detector calculating energy losses and
taking 
into account the detector energy and angular resolution. The source
angular coordinates  on the celestial sphere are chosen randomly from
a uniform distribution in right ascension and with a declination
distribution proportional to $\cos\delta$.\footnote{The declination
  $\delta$ is defined as being $0^\circ$ on the equatorial plane and
  $\pm90^\circ$ at the poles, so 
$\ud \Omega = \cos\delta \ud \delta \ud \alpha$.} 
Since we neglect the effects of the magnetic fields on the
propagation, we ignore sources outside the experimental field of view
and assign to visible sources a probability proportional to: 
\begin{equation}
   r(z)^{-2} \omega(\delta)\,,
\end{equation}
where $r(z)^{-2}$ takes into account the distance dependence of the
solid angle and $\omega(\delta)$ is the relative exposure of the
experiment in the given direction.  

For a given simulated set of events, we calculate the angular  two
point correlation function as defined in \cite{isola,sato2} to study
the expected departure from an isotropic distributions in the
sky. For each event, the number of events within a circle at an
angular distance $\theta$ and a bin width $\Delta \theta$ is summed: 
\begin{equation}
N(\theta) = \frac{1}{S(\theta)}\sum_{i>j} R_{i j}(\theta),
\end{equation} 
and divided by the area $S(\theta)=2\pi |\cos(\theta)-\cos(\theta+\Delta 
\theta)|$ 
of the angular bin between $\theta$ and $\theta+\Delta \theta$,  and 
$R_{i j}(\theta)=0,1$ counts the events in the same bin.

For the Auger observatory in the Southern Hemisphere, we assume a
total acceptance of 7,000 $\rm km^2~sr$ independent of energy above
$10^{19}\eV$. For the angular resolution we used $1^\circ$, for the
energy resolution 20\%, and for the exposure dependence on the arrival
direction we 
used the analytical estimate given in Ref.~\cite{sommers}.  By
comparison, AGASA had an  acceptance of $160~\rm km^2~sr$ and reported
886 events  above $10^{19}$ eV  over a decade with an accumulated
exposure of $1,645~\rm km^2~sr~yr$. If the two experiments had the
same energy calibration,  Auger would accumulate over $\sim 10^4$
events above $10^{19}$ eV in 5 years of operation. However, the first
released Auger spectrum \cite{sommersicrc} suggests that the energy
calibration of AGASA \cite{agasaspec} is systematically high and,
consequently, a lower flux is observed for a fixed energy at Auger.   

We normalize our simulations to the Auger flux above $4 \times 10^{19}$
eV \cite{sommersicrc} (indicated in Table \ref{table1} as $n_{19.6}$:
number of events above $4 \times 10^{19}\eV$) and determine the number
of events with energy above $10^{20}$ eV as a result of our simulations.
These numbers are summarized in Table \ref{table1}, for a continuous
distribution of sources and for different values of the source density.

\begin{table}
  \centering
\caption{\label{table1} Number of events expected with $E > 10^{20}$ eV.}
\vspace{1ex}
\begin{tabular}{c|*{4}{|@{\hspace{1em}}c@{\hspace{1em}}}}
\hline \hline \hline
  \rule{0pt}{4ex}
  $\bar{\rho}$ (Mpc$^{-3}$) & $n_{20}$ / 1yr &
  $n_{20}$ / 5yr & $n_{20}$ / 10yr &
  $n_{20}$ / 15yr\\
  \rule[-2ex]{0pt}{3ex}
  & $n_{19.6}=70$ & $n_{19.6}=350$ & $n_{19.6}=700$ & $n_{19.6}=1050$ \\
\hline \hline \hline
\rule[-2ex]{0pt}{6ex}
cont. & $2.9\pm1.7$ & $14\pm4$ & $29\pm5$ & $43\pm6$\\\hline
\rule[-2ex]{0pt}{6ex}
$10^{-3}$ & $2.5\pm1.3$ & $13\pm3$ & $26\pm6$ & $39\pm8$\\\hline
\rule[-2ex]{0pt}{6ex}
$10^{-4}$ & $2.0\pm1.4$ & $10\pm4$ & $22\pm6$ & $34\pm7$\\\hline
\rule[-2ex]{0pt}{6ex}
$10^{-5}$ & $1.6\pm1.5$ & $7.3\pm3.3$ & $14\pm5$ & $20\pm7$\\\hline
\rule[-2ex]{0pt}{6ex}
$10^{-6}$ & $0.4\pm0.7$ & $2.1\pm1.7$ & $4.2\pm2.9$ & $7.2\pm4.3$\\
\hline \hline \hline
\end{tabular}
\end{table}
                                              
\begin{figure}
  \centering
  \includegraphics[width=0.9\textwidth]{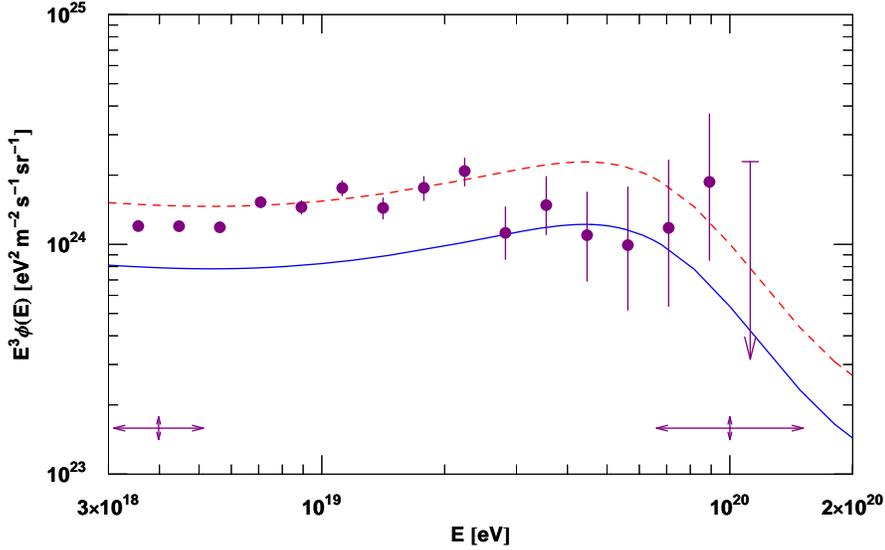}
  \vspace*{-0.3cm}
  \caption{Data points from the Pierre Auger Observatory
  \cite{sommersicrc} compared with theoretical predictions normalized at
  different energies.}\label{fig:spec} 
\end{figure}

The numbers in Table \ref{table1} require some further comments: the
spectrum of cosmic rays as detected by Auger at the present time is
rather odd, in that it appears to have a dip instead of a bump at
energies around $4\times 10^{19}$ eV. This may easily lead to an incorrect 
estimate of the number of events expected at higher energies. In
Fig. \ref{fig:spec} we show the Auger data and the predicted spectrum 
as normalized at low energies ($10^{19}\eV$, dashed line) and at $4\times 10^{19}$
eV (solid line). Systematic uncertainties on the flux and on the energy
determination are indicated by double arrows at two different energies
\cite{sommersicrc}.

The figure clearly illustrates the fact that the normalization at
$4\times 10^{19}$ eV may cause an underestimate of the number of
events expected at higher energies, since most likely the feature
observed at that energy is the result of a statistical fluctuation. 
In order to keep memory of this problem we carry out our analyses for
both instances of normalization. 

The number of events as obtained with the higher normalization are
reported in Table \ref{table2}, an are systematically higher by a factor
$\sim 2$. 

\begin{table}
  \centering
\caption{\label{table2}Number of events expected with $E > 10^{20}$ eV
  using the normalization of the spectrum at low energy.}
\vspace{1ex}
\begin{tabular}{c|*{4}{|@{\hspace{1em}}c@{\hspace{1em}}}}
\hline \hline \hline
  \rule{0pt}{4ex}
  $\bar{\rho}$ (Mpc$^{-3}$) & $n_{20}$ / 1yr &
  $n_{20}$ / 5yr & $n_{20}$ / 10yr &
  $n_{20}$ / 15yr\\
  \rule[-2ex]{0pt}{3ex}
  & $n_{19.6}=120$ & $n_{19.6}=600$ & $n_{19.6}=1200$ & $n_{19.6}=1800$ \\
\hline \hline \hline
\rule[-2ex]{0pt}{6ex}
cont. & $5.0\pm2.2$ & $25\pm5$ & $50\pm7$ & $75\pm8$\\\hline
\rule[-2ex]{0pt}{6ex}
$10^{-3}$ & $4.2\pm1.8$ & $22\pm5$ & $44\pm9$ & $67\pm11$\\\hline
\rule[-2ex]{0pt}{6ex}
$10^{-4}$ & $3.6\pm2.0$ & $18\pm5$ & $38\pm9$ & $58\pm12$\\\hline
\rule[-2ex]{0pt}{6ex}
$10^{-5}$ & $2.5\pm1.8$ & $13\pm5$ & $24\pm8$ & $35\pm10$\\\hline
\rule[-2ex]{0pt}{6ex}
$10^{-6}$ & $0.8\pm1.0$ & $3.5\pm2.3$ & $7.4\pm4.2$ & $13\pm7$\\
\hline \hline \hline
\end{tabular}
\end{table}

We parametrize the mean density for the distribution of sources by
$\bar{\rho} = 10^{-r}$ Mpc$^{-3}$,  and choose  $r$ between 3 and
6. The mean densities of extragalactic astrophysical accelerators of
UHECRs should be well represented by this range which covers galaxies,
groups and clusters of galaxies. For example, black holes in centers
of normal galaxies should have $r \simeq 3$ while those in active
galaxies should have $r \ga 6$. Similar densities are expected for
rich clusters of galaxies. The most common extragalactic objects that may
house UHECR sources are galaxies, which have  $\bar{\rho}$ depending
on their type and luminosity ranging from $r=$ 2 to 3. For example,
the Sloan Digital Sky survey recently reported a comprehensive  study
\cite{SDSS06} of galaxy clustering in a large volume limited sample at
relatively low redshifts ($z$ between 0.015 and 0.1).  In their
samples, galaxy number densities vary between $2 \times 10^{-2} h^3$
Mpc$^{-3}$ (in their  sample with limiting r-magnitude -18)  to  $6
\times 10^{-3} h^3$ Mpc$^{-3}$ (for the brighter sample). The number
density in groups of galaxies range from  $6 \times 10^{-4}h^3$
Mpc$^{-3}$ (starting with groups of 3 galaxies) to richer clusters
with $10^{-7}h^3$ Mpc$^{-3}$. Choosing  the Hubble parameter, $h =
H_0/100$ km/s/Mpc = 0.75,  the observed $\bar{\rho}$ range from $8
\times 10^{-3}$ Mpc$^{-3}$ to $4 \times 10^{-8}$  Mpc$^{-3}$. As we
discuss below, detecting SSAs for $\bar{\rho} \ga 10^{-3}$ Mpc$^{-3}$
will require many years of observations, while $\bar{\rho} \ll
10^{-6}$ Mpc$^{-3}$ may generate higher UHECR clustering than reported
so far, unless magnetic fields move the threshold for charged
particle astronomy to energies around $10^{20}\eV$. Such low densities
of sources however would generate a sharp GZK cutoff in the diffuse
spectrum, starting at energies even below $10^{20}$ eV. This situation
appears to be disfavored even on the basis of current data.

\subsection{Two point correlation function above $4 \times 10^{19}\eV$ }

\begin{figure}
  \centering
  \includegraphics[width=0.9\textwidth]{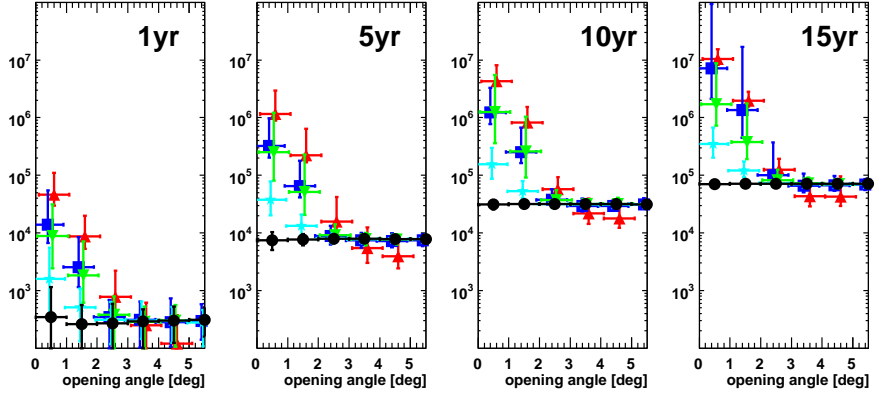}
  \vspace*{-0.3cm}
  \caption{Two point correlation function expected for Auger events
  above $4 \times 10^{19}\eV$ after 1, 5, 10, and 15 years, for
  continuous distribution of sources (black circles),  and number
  densities of $10^{-3}$ Mpc$^{-3}$ (cyan stars), $10^{-4}$ Mpc$^{-3}$
  (green downward triangles),  $10^{-5}$ Mpc$^{-3}$ (blue squares),
  and $10^{-6}$ Mpc$^{-3}$ (red upward triangles).  No limitation on
  the minimum distance for sources were imposed. Error bars are
  asymmetric 1 $\sigma$. The data points in this and in all subsequent
  figures have been slightly offset (horizontally) for clarity of
  presentation.}\label{fig:auger_2pcf_4e19_0} 
\end{figure}

Figure \ref{fig:auger_2pcf_4e19_0} shows the two point correlation
function of simulated events with energies above  $4 \times
10^{19}\eV$ for  the full Auger South aperture and exposures after 1, 
5, 10, and 15 years of operation. The number of events is normalized
at the Auger data at energy $4\times 10^{19}$ eV. The black circles show a
continuous distribution of sources, while sources with number
densities of $10^{-3}$ Mpc$^{-3}$ are represented by cyan stars,
$10^{-4}$ Mpc$^{-3}$ by green downward triangles,  $10^{-5}$
Mpc$^{-3}$ by blue squares, and $10^{-6}$ Mpc$^{-3}$ by red upward
triangles. The error bars are asymmetric  1 $\sigma$ standard deviations
calculated in each energy bin for events above and below the mean.
It is clear from this figure that after one year with the full Auger
South aperture a discrete distribution with $\bar{\rho} \la 10^{-4}$
Mpc$^{-3}$  should be distinguishable from a continuous source
distribution. After 5 years, even a  $\bar{\rho} \sim 10^{-3}$
Mpc$^{-3}$ can be detected at the few $\sigma$ level. After a decade,
departure from continuous should be detected at many tens of
$\sigma$. 
 
Although departures from  continuous distributions should become clear
soon, distinguishing between different specific source densities in
Fig. \ref{fig:auger_2pcf_4e19_0} is more difficult.  

\begin{figure}
  \centering
  \vspace*{-0.2cm}
  \includegraphics[width=0.9\textwidth]{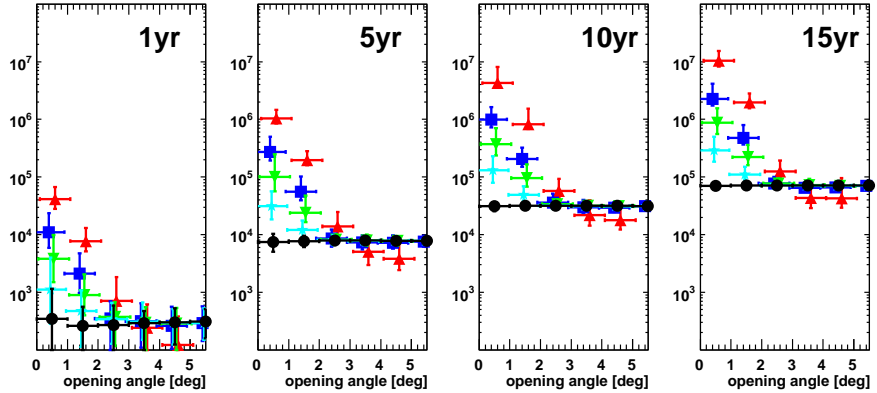}
  \vspace*{-0.3cm}
  \caption{Same as Fig.\ref{fig:auger_2pcf_4e19_0} but the minimum
  distance for nearby sources is $d_\mathrm{min}(\bar{\rho})$. 
  }\label{fig:auger_2pcf_4e19_var}
\end{figure}

In the 100 realizations for the assumptions in
Fig. \ref{fig:auger_2pcf_4e19_0}, there are often configurations with
sources within 10 Mpc. These randomly placed nearby sources produce a
significant dispersion in the two point correlation function at small
opening angles and generate an overlap between different  choices of
$\bar{\rho}$. To help distinguish  between different number densities,
we  impose a lower  limit in the distance to the nearest source and
show the constrained angular correlations in  Fig.
\ref{fig:auger_2pcf_4e19_var}. In Fig. \ref{fig:auger_2pcf_4e19_var},
the nearest source is at a distance larger than
$d_{min}(\bar{\rho})$, where  
\begin{equation}
   d_{min}(\bar{\rho}) =  {1\over 2}  \ \bar{\rho}^{-1/3}  \ .
\end{equation}
As can be seen in  Fig.~\ref{fig:auger_2pcf_4e19_var}, imposing a minimum
distance alleviates  the degeneracies between different source
densities after 5 years of full Auger South operations. After the
first full year exposure, number densities above  $10^{-4}$ Mpc$^{-3}$
give clear departure from continuous distributions,  while for
$\la10^{-3}$ Mpc$^{-3}$, 5 years are necessary. 

In Table \ref{table3}, we list the mean number of doublets
($\bar{N}_\mathrm{d}$) expected after 1, 5, 10, and 15 years of Auger
South operations for $E > 4 \times 10^{19}$ eV and for $r$ between 3
and 6. In the same table we report, along with $\bar{N}_\mathrm{d}$, the
asymmetric $1\sigma$ errorbars for realizations with a number of
doublets above or below the mean.
The opening angle used to count doublets is 1$^\circ$
and the realizations have minimum source distance
$d_\mathrm{min}(\bar{\rho})$ as in Fig. \ref{fig:auger_2pcf_4e19_var}. 
It is worth stressing that $\bar{N}_\mathrm{d}$ in the table is
the total number of doublets, including doublets inside possible
higher multiplicity clusters.  

\begin{table}
  \centering
\caption{\label{table3}Number of doublets for  $E > 4 \times
10^{19}$ eV, using the normalization of the number of simulated events
to the number of events observed by Auger above $4\times 10^{19}$ eV.}
\vspace{1ex}
\begin{tabular}{c|*{4}{|@{\hspace{1em}}c@{\hspace{1em}}}}
\hline \hline \hline
  \rule[-2ex]{0pt}{6ex}
  $\bar{\rho}$ (Mpc$^{-3}$) & $\bar{N}_\mathrm{d}$ / 1yr &
  $\bar{N}_\mathrm{d}$ / 5yr & $\bar{N}_\mathrm{d}$ / 10yr &
  $\bar{N}_\mathrm{d}$ / 15yr\\
\hline \hline \hline
\rule[-2ex]{0pt}{6ex}
cont. & $0.33_{\rule{0pt}{6pt}-0.33}^{\rule[-3pt]{0pt}{3pt}+0.77}$ &
$7.13_{\rule{0pt}{6pt}-2.33}^{\rule[-3pt]{0pt}{3pt}+2.77}$ &
$29.7_{\rule{0pt}{6pt}-5.3}^{\rule[-3pt]{0pt}{3pt}+5.3}$ &
$67.5_{\rule{0pt}{6pt}-8.3}^{\rule[-3pt]{0pt}{3pt}+8.1}$\\\hline
\rule[-2ex]{0pt}{6ex}
$10^{-3}$ & $1.07_{\rule{0pt}{6pt}-0.80}^{\rule[-3pt]{0pt}{3pt}+2.75}$ &
$29.7_{\rule{0pt}{6pt}-12.0}^{\rule[-3pt]{0pt}{3pt}+24.3}$ &
$125_{\rule{0pt}{6pt}-49}^{\rule[-3pt]{0pt}{3pt}+94}$ &
$279_{\rule{0pt}{6pt}-104}^{\rule[-3pt]{0pt}{3pt}+201}$\\\hline
\rule[-2ex]{0pt}{6ex}
$10^{-4}$ & $3.63_{\rule{0pt}{6pt}-2.21}^{\rule[-3pt]{0pt}{3pt}+5.84}$ &
$96.7_{\rule{0pt}{6pt}-42.3}^{\rule[-3pt]{0pt}{3pt}+139.1}$ &
$355_{\rule{0pt}{6pt}-129}^{\rule[-3pt]{0pt}{3pt}+318}$ &
$844_{\rule{0pt}{6pt}-313}^{\rule[-3pt]{0pt}{3pt}+647}$\\\hline
\rule[-2ex]{0pt}{6ex}
$10^{-5}$ & $10.6_{\rule{0pt}{6pt}-5.0}^{\rule[-3pt]{0pt}{3pt}+12.0}$ &
$259_{\rule{0pt}{6pt}-75}^{\rule[-3pt]{0pt}{3pt}+217}$ &
$953_{\rule{0pt}{6pt}-257}^{\rule[-3pt]{0pt}{3pt}+619}$ &
$2,181_{\rule{0pt}{6pt}-530}^{\rule[-3pt]{0pt}{3pt}+1,763}$\\\hline
\rule[-2ex]{0pt}{6ex}
$10^{-6}$ & $39.6_{\rule{0pt}{6pt}-12.8}^{\rule[-3pt]{0pt}{3pt}+24.6}$ &
$987_{\rule{0pt}{6pt}-173}^{\rule[-3pt]{0pt}{3pt}+408}$ &
$4,078_{\rule{0pt}{6pt}-658}^{\rule[-3pt]{0pt}{3pt}+3,702}$ &
$9,946_{\rule{0pt}{6pt}-2,010}^{\rule[-3pt]{0pt}{3pt}+4,681}$\\
\hline \hline \hline
\end{tabular}
\end{table}

If the number of events used in the simulation is normalized to the low
energy spectrum as measured by Auger South, then the number of
doublets expected in the future Auger operation are those reported in
Table \ref{table4}.

\begin{table}
  \centering
\caption{\label{table4}Number of doublets for  $E > 4 \times
10^{19}$ eV, using the normalization of the number of simulated events
to the number of events observed by Auger at low energy.}
\vspace{1ex}
\begin{tabular}{c|*{4}{|@{\hspace{1em}}c@{\hspace{1em}}}}
\hline \hline \hline
  \rule[-2ex]{0pt}{6ex}
  $\bar{\rho}$ (Mpc$^{-3}$) & $\bar{N}_\mathrm{d}$ / 1yr &
  $\bar{N}_\mathrm{d}$ / 5yr & $\bar{N}_\mathrm{d}$ / 10yr &
  $\bar{N}_\mathrm{d}$ / 15yr\\
\hline \hline \hline
\rule[-2ex]{0pt}{6ex}
cont. & $0.92_{\rule{0pt}{6pt}-0.92}^{\rule[-3pt]{0pt}{3pt}+0.99}$ &
$22.0_{\rule{0pt}{6pt}-4.6}^{\rule[-3pt]{0pt}{3pt}+4.8}$ &
$88.7_{\rule{0pt}{6pt}-9.7}^{\rule[-3pt]{0pt}{3pt}+9.1}$ &
$200_{\rule{0pt}{6pt}-12}^{\rule[-3pt]{0pt}{3pt}+16}$\\\hline
\rule[-2ex]{0pt}{6ex}
$10^{-3}$ & $3.36_{\rule{0pt}{6pt}-2.03}^{\rule[-3pt]{0pt}{3pt}+4.51}$ &
$91.0_{\rule{0pt}{6pt}-35.3}^{\rule[-3pt]{0pt}{3pt}+73.0}$ &
$366_{\rule{0pt}{6pt}-109}^{\rule[-3pt]{0pt}{3pt}+300}$ &
$830_{\rule{0pt}{6pt}-254}^{\rule[-3pt]{0pt}{3pt}+562}$\\\hline
\rule[-2ex]{0pt}{6ex}
$10^{-4}$ & $11.0_{\rule{0pt}{6pt}-5.6}^{\rule[-3pt]{0pt}{3pt}+19.6}$ &
$269_{\rule{0pt}{6pt}-101}^{\rule[-3pt]{0pt}{3pt}+338}$ &
$1,031_{\rule{0pt}{6pt}-373}^{\rule[-3pt]{0pt}{3pt}+821}$ &
$2,448_{\rule{0pt}{6pt}-874}^{\rule[-3pt]{0pt}{3pt}+1,754}$\\\hline
\rule[-2ex]{0pt}{6ex}
$10^{-5}$ & $31.3_{\rule{0pt}{6pt}-12.6}^{\rule[-3pt]{0pt}{3pt}+35.1}$ &
$773_{\rule{0pt}{6pt}-222}^{\rule[-3pt]{0pt}{3pt}+664}$ &
$2,814_{\rule{0pt}{6pt}-757}^{\rule[-3pt]{0pt}{3pt}+1,831}$ &
$6,457_{\rule{0pt}{6pt}-1,540}^{\rule[-3pt]{0pt}{3pt}+5,427}$\\\hline
\rule[-2ex]{0pt}{6ex}
$10^{-6}$ & $114_{\rule{0pt}{6pt}-27}^{\rule[-3pt]{0pt}{3pt}+46}$ &
$2,884_{\rule{0pt}{6pt}-419}^{\rule[-3pt]{0pt}{3pt}+1,315}$ &
$11,914_{\rule{0pt}{6pt}-1,815}^{\rule[-3pt]{0pt}{3pt}+10,889}$ &
$29,369_{\rule{0pt}{6pt}-5,831}^{\rule[-3pt]{0pt}{3pt}+13,356}$\\
\hline \hline \hline
\end{tabular}
\end{table}
                                                                               
\subsection{Two point correlation function  above $10^{20} eV$ }

UHECR clustering and source positions should become easier to identify
with events of energies  $\ga 10^{20}\eV$, given that at these
energies UHECRs are most likely protons and their trajectories are
less likely to be affected by extragalactic and galactic magnetic fields 
than lower
energy events.   The difficulty is clearly in the limited statistics of events
that can be accumulated in this energy range by Auger South, since the
flux is a steeply decreasing power of energy and energy losses that
give rise to the GZK feature are also at play in this energy range. A
larger array as currently being discussed for the Auger North
Observatory should help significantly the ability to do charged
particle astronomy at this energy scale. 

\begin{figure}
  \centering
  \includegraphics[width=0.9\textwidth]{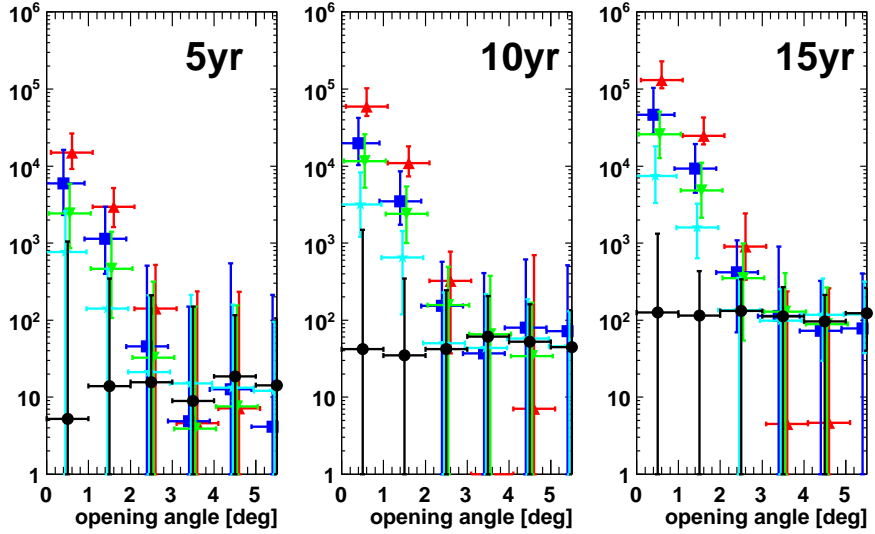}
  \vspace*{-0.3cm}
  \caption{Two point correlation function for events above
  $10^{20}\eV$ after 5, 10, and 15 years at   Auger South for a
  continuous distribution of sources (black circles),  and number
  densities of  $10^{-3}$ Mpc$^{-3}$ (cyan stars), $10^{-4}$
  Mpc$^{-3}$ (green downward triangles),  $10^{-5}$ Mpc$^{-3}$ (blue
  squares), and $10^{-6}$ Mpc$^{-3}$ (red upward triangles).  The
  minimum distance for the nearest source is  $d_\mathrm{min}(\bar{\rho})$.  
  }\label{fig:auger_2pcf_1e20_var_fixed}
\end{figure}

We simulated the expected angular correlation function in Auger South
for events with energies above $10^{20}$ eV for different source
densities. In Fig. \ref{fig:auger_2pcf_1e20_var_fixed},  
the correlation function is shown for sources located at minimum
distances $d_\mathrm{min}(\bar{\rho})$ as in Fig.
\ref{fig:auger_2pcf_4e19_var}. In addition, we fixed the number of
events above  $10^{20}$ eV to the mean number given in Table
\ref{table1}. If we let the number of events fluctuate between
realizations, the error bars are significantly enhanced.  In
principle, once Auger has run for 5 years, we will know the number of
events above  $10^{20}\eV$.  As discussed below, the number of events
expected above $10^{20}\eV$ for a fixed injection spectrum, depends on
the source density. Combining a study of the GZK feature with the
small scale anisotropies should yield $\bar{\rho}$. 
As shown in Fig.  \ref{fig:auger_2pcf_4e19_var}, a departure from a
continuous source distribution is detectable at the 1 $\sigma$ level
after 5 years for $\bar{\rho} \la 10^{-5}$ Mpc$^{-3}$. Larger
densities require 10 to 15 years before clear detectability. Even
after a decade, distinguishing between different source densities will
be a major challenge. 

In Table \ref{table5} we show the predicted number of doublets at 
energies above $10^{20}\eV$ after 5, 10, and 15 years of operation of
Auger south for different values of the source density.

\begin{table}
  \centering
\caption{\label{table5}Number of doublets for  $E > 10^{20}$ eV with
  normalization to the Auger data above $4\times 10^{19}$ eV.}
\vspace{1ex}
\begin{tabular}{c|*{3}{|@{\hspace{1em}}c@{\hspace{1em}}}}
\hline \hline \hline
  \rule[-2ex]{0pt}{6ex}
  $\bar{\rho}$ (Mpc$^{-3}$) &
  $\bar{N}_\mathrm{d}$ / 5yr & $\bar{N}_\mathrm{d}$ / 10yr &
  $\bar{N}_\mathrm{d}$ / 15yr\\
\hline \hline \hline
\rule[-2ex]{0pt}{6ex}
cont. & $0.02_{\rule{0pt}{6pt}-0.02}^{\rule[-3pt]{0pt}{3pt}+1.40}$ & 
$0.04_{\rule{0pt}{6pt}-0.04}^{\rule[-3pt]{0pt}{3pt}+1.38}$ &
$0.12_{\rule{0pt}{6pt}-0.12}^{\rule[-3pt]{0pt}{3pt}+1.15}$\\\hline
\rule[-2ex]{0pt}{6ex}
$10^{-3}$ & $0.73_{\rule{0pt}{6pt}-0.73}^{\rule[-3pt]{0pt}{3pt}+1.76}$ &
$3.05_{\rule{0pt}{6pt}-1.90}^{\rule[-3pt]{0pt}{3pt}+4.84}$ &
$7.09_{\rule{0pt}{6pt}-3.91}^{\rule[-3pt]{0pt}{3pt}+10.11}$\\\hline
\rule[-2ex]{0pt}{6ex}
$10^{-4}$ & $2.32_{\rule{0pt}{6pt}-1.50}^{\rule[-3pt]{0pt}{3pt}+3.28}$ &
$11.1_{\rule{0pt}{6pt}-6.1}^{\rule[-3pt]{0pt}{3pt}+13.7}$ &
$24.8_{\rule{0pt}{6pt}-12.6}^{\rule[-3pt]{0pt}{3pt}+23.9}$\\\hline
\rule[-2ex]{0pt}{6ex}
$10^{-5}$ & $5.71_{\rule{0pt}{6pt}-3.48}^{\rule[-3pt]{0pt}{3pt}+9.77}$ &
$19.0_{\rule{0pt}{6pt}-9.1}^{\rule[-3pt]{0pt}{3pt}+21.6}$ &
$44.4_{\rule{0pt}{6pt}-19.4}^{\rule[-3pt]{0pt}{3pt}+54.6}$\\\hline
\rule[-2ex]{0pt}{6ex}
$10^{-6}$ & $14.3_{\rule{0pt}{6pt}-5.6}^{\rule[-3pt]{0pt}{3pt}+11.0}$ &
$56.8_{\rule{0pt}{6pt}-14.2}^{\rule[-3pt]{0pt}{3pt}+41.6}$ &
$125_{\rule{0pt}{6pt}-27}^{\rule[-3pt]{0pt}{3pt}+94}$\\
\hline \hline \hline
\end{tabular}
\end{table}

\begin{figure}
  \centering
  \includegraphics[width=0.9\textwidth]{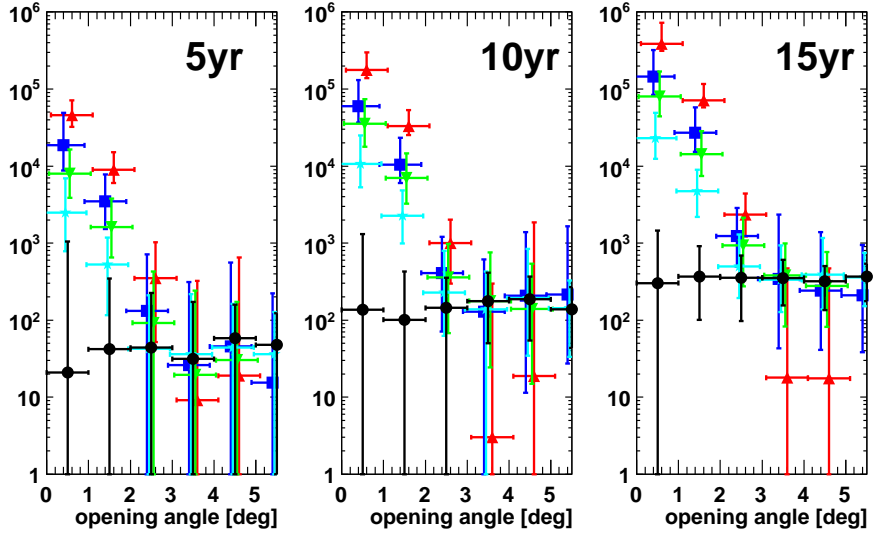}
  \vspace*{-0.3cm}
  \caption{Two point correlation function for events above
  $10^{20}\eV$ after 5, 10, and 15 years at   Auger South for a
  continuous distribution of sources (black circles),  and number
  densities of  $10^{-3}$ Mpc$^{-3}$ (cyan stars), $10^{-4}$
  Mpc$^{-3}$ (green downward triangles),  $10^{-5}$ Mpc$^{-3}$ (blue
  squares), and $10^{-6}$ Mpc$^{-3}$ (red upward triangles).  The
  minimum distance for the nearest source is  $d_\mathrm{min}(\bar{\rho})$.  
  }\label{fig:2pcfnew}
\end{figure}

As stressed in the previous section, the number of events at energies
above $10^{20}$ eV is estimated by using as a normalization the flux
currently observed by Auger South at $4\times 10^{19}$ eV. However, as 
shown in Fig. \ref{fig:spec}, the observed spectrum appears to be at odds 
with the observed low energy spectrum. If the solid line in
Fig. \ref{fig:spec} is used for the normalization of the number 
of events above $10^{20}$ eV, then the situation improves somewhat
(the number of events above $10^{20}$ eV is roughly doubled in this
case). This is shown in Fig. \ref{fig:2pcfnew}, where we plot the 
two point correlation function for the new number of events above 
$10^{20}\eV$ after 5, 10, and 15 years.

The number of doublets obtained for this second instance of
normalization to the data is shown in Table \ref{table6}.

\begin{table}
  \centering
\caption{\label{table6}Number of doublets for  $E > 10^{20}$ eV with
  normalization to the low energy Auger data.}
\vspace{1ex}
\begin{tabular}{c|*{3}{|@{\hspace{1em}}c@{\hspace{1em}}}}
\hline \hline \hline
  \rule[-2ex]{0pt}{6ex}
  $\bar{\rho}$ (Mpc$^{-3}$) &
  $\bar{N}_\mathrm{d}$ / 5yr & $\bar{N}_\mathrm{d}$ / 10yr &
  $\bar{N}_\mathrm{d}$ / 15yr\\
\hline \hline \hline
\rule[-2ex]{0pt}{6ex}
cont. & $0.02_{\rule{0pt}{6pt}-0.02}^{\rule[-3pt]{0pt}{3pt}+0.98}$ &
$0.19_{\rule{0pt}{6pt}-0.19}^{\rule[-3pt]{0pt}{3pt}+0.98}$ &
$0.39_{\rule{0pt}{6pt}-0.39}^{\rule[-3pt]{0pt}{3pt}+1.05}$\\\hline
\rule[-2ex]{0pt}{6ex}
$10^{-3}$ & $3.29_{\rule{0pt}{6pt}-2.08}^{\rule[-3pt]{0pt}{3pt}+5.08}$ &
$13.9_{\rule{0pt}{6pt}-7.1}^{\rule[-3pt]{0pt}{3pt}+15.4}$ &
$30.0_{\rule{0pt}{6pt}-13.8}^{\rule[-3pt]{0pt}{3pt}+32.6}$\\\hline
\rule[-2ex]{0pt}{6ex}
$10^{-4}$ & $10.5_{\rule{0pt}{6pt}-5.3}^{\rule[-3pt]{0pt}{3pt}+9.5}$ &
$46.8_{\rule{0pt}{6pt}-22.8}^{\rule[-3pt]{0pt}{3pt}+50.1}$ &
$102_{\rule{0pt}{6pt}-46}^{\rule[-3pt]{0pt}{3pt}+104}$\\\hline
\rule[-2ex]{0pt}{6ex}
$10^{-5}$ & $23.9_{\rule{0pt}{6pt}-12.5}^{\rule[-3pt]{0pt}{3pt}+33.4}$ &
$76.8_{\rule{0pt}{6pt}-29.4}^{\rule[-3pt]{0pt}{3pt}+94.3}$ &
$187_{\rule{0pt}{6pt}-78}^{\rule[-3pt]{0pt}{3pt}+220}$\\\hline
\rule[-2ex]{0pt}{6ex}
$10^{-6}$ & $59.6_{\rule{0pt}{6pt}-17.9}^{\rule[-3pt]{0pt}{3pt}+33.9}$ &
$228_{\rule{0pt}{6pt}-46}^{\rule[-3pt]{0pt}{3pt}+168}$ &
$503_{\rule{0pt}{6pt}-93}^{\rule[-3pt]{0pt}{3pt}+413}$\\
\hline \hline \hline
\end{tabular}
\end{table}

\subsection{Spectra for different number densities}

As shown in \cite{DBO05} the combination of small angle clustering and
spectral informations is a powerful tool to investigate the nature of
the sources of UHECRs. A low density of sources would in fact be
responsible for more evident small angle clustering of the arrival
directions, but at the same time would also induce a sharper shape of
the GZK feature, and therefore a lower number of events at the highest
energies. A larger density of sources has the opposite effect: the GZK
feature is shallower (more events at the highest energies) but the
clustering is mild. As pointed out in \cite{DBO05}, the AGASA data on
the small scale anisotropies appear to be at odds with the spectrum,
as measured by the same experiment. 

Here we investigate the potential to use a similar technique for the
data of the Pierre Auger Observatory. 

\begin{figure}
  \centering
  \includegraphics[width=0.9\textwidth]{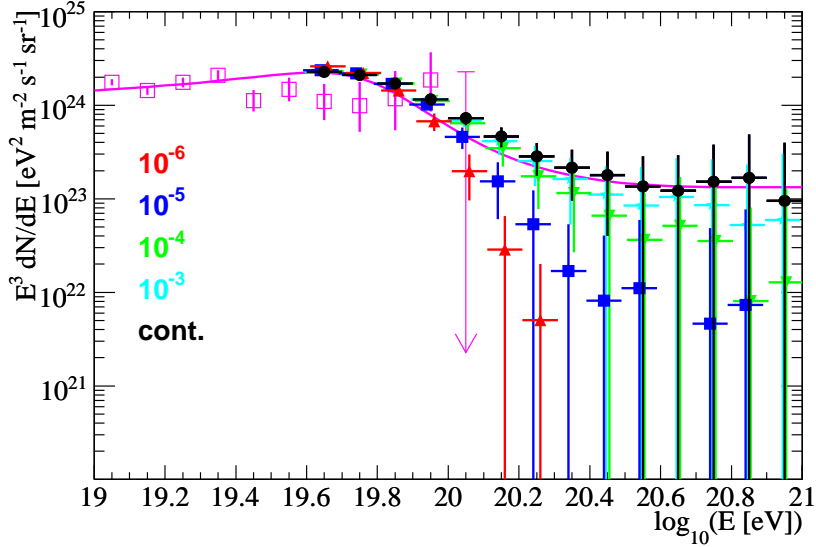}
  \caption{Energy spectrum for different source densities compared 
  with Auger data (open squares).
  The sources densities are $10^{-6}$ Mpc$^{-3}$ (red upward
  triangles); 
  $10^{-5}$ Mpc$^{-3}$ (blue
  squares),  $10^{-4}$ Mpc$^{-3}$ (green downward triangles);
  $10^{-3}$ Mpc$^{-3}$ (cyan stars); and a continuous distribution of
  sources (black circles).} 
   \label{fig:Auger_15yr}
\end{figure}

In Fig. \ref{fig:Auger_15yr}, we show the energy spectrum around the
GZK feature that may be detected by Auger South after 15 years of full
aperture operations, assuming the normalization at low energy. The
spectrum is shown for different source densities ranging from a
continuous source distribution (black circles) to source densities
between $10^{-3}$ Mpc$^{-3}$ (cyan stars) and  $10^{-6}$ Mpc$^{-3}$ (red
upward triangles). Also shown is the first release of the Auger spectrum
(open magenta squares) \cite{sommersicrc} and an analytical prediction
(continuous line). 

The number of events at energy above $10^{20}$ eV is a rather
sensitive function of the source density, as can also be understood
from Tables \ref{table1} and \ref{table2}. For the case of
normalization at low energy, one could have as many as $\sim 75$
events above $10^{20}$ eV in 15 years of operation of Auger South, 
for a continuous (and unlikely) distribution of sources, or as little 
as $\sim 13$ events in the same energy region if the sources have a 
mean local density $10^{-6}\rm Mpc^{-3}$. Unfortunately this also
means that the statistical uncertainties in the fluxes at the highest
energies become large if the source density is too low, and it becomes
correspondingly harder to have a precision measurement of the shape of
the GZK feature in these cases. For source densities between
$10^{-6}\rm Mpc^{-3}$ and  $10^{-5}\rm Mpc^{-3}$ Auger is expected to
become statistics dominated at energies $\sim 10^{20.2}$
eV. However at this energy the different predicted spectra already
differ from that generated by a continuous source distribution by about
a factor $\sim 10$. Again, the combination of spectra measurements and
anisotropy measurements can play an instrumental role in inferring
hints to the nature of the sources of UHECRs. 

\section{Conclusions}

The imminent completion of the Auger Observatory in Argentina will
mark the beginning of a new era in cosmic ray astrophysics. The
combination of a very large ground array and fluorescence telescopes
will provide a detailed measurement of the spectrum at the highest
energies. 

One of the main open questions in Cosmic Ray Physics is the presence 
of a GZK feature in the spectrum. If this feature is in fact there, it
becomes very important to measure its shape, which can provide us with
information on the number of sources and their spatial distribution. 
Independently of the presence or absence of the GZK feature, the
problem of finding an acceleration mechanism and a class of sources 
that may harbor the accelerator remains of paramount importance. Auger 
will most likely clarify whether the spectrum of UHECRs has or not the GZK
suppression. In order to answer the second big
question, we need high quality information (namely high statistics) 
about the directions of arrival of UHECRs, and their chemical
composition. The first hints of such new era of charged particle
astronomy are likely to come from small deviations from isotropic 
distributions in the next few years of Auger South operations. 

In this paper we made quantitative predictions of the amount of small 
scale anisotropies expected in Auger in the next decade for a range 
of plausible source densities. We investigated two instances of
normalization of the number of events expected at the highest
energies, depending on whether the procedure is applied to the data
measured by Auger at $4\times 10^{19}$ eV or rather at lower energies
where statistical errors should be less important. The amount of small
angle clustering has been quantified through the two point correlation
function, both for events above $4\times 10^{19}$ eV and above
$10^{20}$ eV. We conclude that it will be sufficient to run Auger for
a few years to detect the departure of the two point correlation
function from that expected from a continuous distribution of
sources. On the other hand, it will be more problematic to measure the
mean source density with high accuracy, because of the relatively low
number of events and correspondingly large statistical fluctuations. As already
proposed in \cite{DBO05}, the combination of small angle clustering
and spectrum of diffuse cosmic rays is able to provide us with more
stringent constraints on the source density. This will be true for
Auger as well, although the statistical error bars on the spectrum for
the case of discrete sources, as shown in Fig. \ref{fig:Auger_15yr},
are expected to become relatively large at energies around $10^{20.2}$ eV. 
This result is rather more pessimistic than the ones previously
obtained in \cite{DBO03} because such previous results were obtained
adopting the normalization (not the shape) of the AGASA spectrum,
which has now been found to lead to cosmic ray fluxes about three
times larger than those actually measured by Auger at the present
time. 

Despite these problems, it is fair to conclude that once the
departures from isotropy will be detected, the estimate of the mean
density of sources, $\bar \rho$ will be within reach. 
The even more challenging possibility of measuring the spectrum of a
single source \cite{BD04} will require a significant increase in 
statistics at the highest energies as currently being discussed in 
the context of the Auger North Observatory or future generation 
cosmic ray observatories, that will most likely be located in space. 

\ack 
The work of D.D.M. is funded through NASA APT grant ATP03-0000-0080 at
University of Delaware. The research of P.B. is funded through
COFIN2004/2005. This work was supported in part by the KICP under NSF
PHY-0114422 and by NSF PHY-0457069, at the University of Chicago.

\end{document}